\documentclass[a4paper, conference]{IEEEtran} 
\usepackage{amsmath,amssymb}
\usepackage{epsfig}
\usepackage{times}
\usepackage{tabularx}
\usepackage{color}
\usepackage{url}
\usepackage{epstopdf}
\usepackage[lined,linesnumbered,titlenumbered,ruled,noend]{algorithm2e}
\usepackage{bm}
\usepackage{mathtools}
\usepackage{booktabs}
\usepackage[absolute]{textpos}
\begin{document}

\begin{textblock}{15}(3,1)
\centering\small A reduced version of this manuscript has been accepted to the IEEE/ACM-IWQoS 2018
\end{textblock}
%\begin{textblock}{5}(1,27)
%\includegraphics[height=1.cm,width=5cm]{Demo}
%\end{textblock}

\title{SWAM: SDN-based Wi-Fi Small Cells with Joint Access-Backhaul and  Multi-Tenant Capabilities}

\author{
    \IEEEauthorblockN{Matteo Grandi\IEEEauthorrefmark{1}, Daniel Camps-Mur\IEEEauthorrefmark{1}, August Betzler\IEEEauthorrefmark{1}, Joan Josep Aleixendri\IEEEauthorrefmark{1}, and Miguel Catalan-Cid\IEEEauthorrefmark{1}}
    \IEEEauthorblockA{\IEEEauthorrefmark{1}i2CAT Foundation
    \\\{matteo.grandi, daniel.camps, august.betzler, joan.aleixendri, miguel.catalan\}@i2cat.net
		}
     %This manuscript has been reviewed and accepted as reduced version to the IEEE/ACM-IWQoS 2018 
%International Symposium on Quality of Service.
}

\maketitle

% \author{Matteo Grandi}
% \affiliation{%
%   \institution{i2CAT Foundation}
%   \city{Barcelona} 
%   \postcode{08034}
% }
% \email{matteo.grandi@i2cat.net}

% \author{Daniel Camps-Mur}
% \affiliation{%
%   \institution{i2CAT Foundation}
%   \city{Barcelona} 
%   \postcode{08034}
% }
% \email{daniel.camps@i2cat.net}

% \author{August Betzler}
% \affiliation{%
%   \institution{i2CAT Foundation}
%   \city{Barcelona} 
%   \postcode{08034}
% }
% \email{august.betzler@i2cat.net}

% \author{Joan Josep Aleixendri}
% \affiliation{%
%   \institution{i2CAT Foundation}
%   \city{Barcelona} 
%   \postcode{08034}
% }
% \email{jj.aleixendri@i2cat.net}

% \author{Miguel Catalan-Cid}
% \affiliation{%
%   \institution{i2CAT Foundation}
%   \city{Barcelona} 
%   \postcode{08034}
% }
% \email{miguel.catalan@i2cat.net}

%\renewcommand{\shortauthors}{Camps-Mur et al.}

%\author{ \parbox{3 in}{\centering Huibert Kwakernaak*
%         \thanks{*Use the $\backslash$thanks command to put information here}\\
%         Faculty of Electrical Engineering, Mathematics and Computer Science\\
%         University of Twente\\
%         7500 AE Enschede, The Netherlands\\
%         {\tt\small h.kwakernaak@autsubmit.com}}
%         \hspace*{ 0.5 in}
%         \parbox{3 in}{ \centering Pradeep Misra**
%         \thanks{**The footnote marks may be inserted manually}\\
%        Department of Electrical Engineering \\
%         Wright State University\\
%         Dayton, OH 45435, USA\\
%         {\tt\small pmisra@cs.wright.edu}}
%}

\begin{abstract}
Dense deployments of Small Cells are required to deliver the capacity promised by 5G networks. In this paper we present SWAM, a system that builds on commodity Wi-Fi routers with multiple wireless interfaces to provide a wireless access infrastructure supporting multi-tenancy, mobility, and integrated wireless access and backhaul. An infrastructure provider can deploy inexpensive SWAM nodes to cover a given geographical area, and re-sell this capacity to provide on-demand connectivity for Mobile Network Operators. Our main contribution is the design of the SWAM datapath and control plane, which are inspired by the overlay techniques used to enable multi-tenancy in data-center networks. We prototype SWAM in an office wireless testbed, and validate experimentally its functionality.
\end{abstract}


%\ccsdesc[500]{Computer systems organization~Embedded systems}

%\keywords{Small Cells, Wireless Backhaul, Multi-tenancy, Slicing}

\maketitle

\section{Introduction}

Densifying the wireless access is seen as the most promising approach to deliver the capacities required in the future 5G network \cite{5g_densification}. Densification can be achieved by rolling out outdoor Small Cells (SCs), which complement the coverage offered by the macro-cell layer with additional capacity in targeted areas. However, three main problems need to be solved to enable massive deployments of outdoor SCs. First, dense SC deployments are prone to interference, which should be mitigated for a maximum performance. Second, SCs will be deployed on street furniture, where network connectivity is often not available at competitive pricing, hence innovative backhaul solutions are needed. Third, being locations for outdoor SCs limited, operators should be able to efficiently share SC infrastructure

A solution to mitigate interference is to aggregate licensed and unlicensed bands. In this regard, the 3GPP has already developed mechanisms that allow to integrate unlicensed spectrum into the mobile network, either by deploying cellular technology in the unlicensed band, or by tightly integrating cellular and Wi-Fi technologies \cite{cellular_unlicensed}.

To address the lack of fiber availability, wireless backhauling is seen as a key technology. %[TODO: Miguel advices to place a reference here to reinforce that wireless BH is a key tech] 
Backhauling based on high capacity mmWave radios is preferable, however cluttered city environments may not always allow for line of sight operation, which forces operators to also consider technologies operating at lower frequencies, below 6 GHz. 

\begin{figure}
	\centering		
        \includegraphics[width=0.5\textwidth]{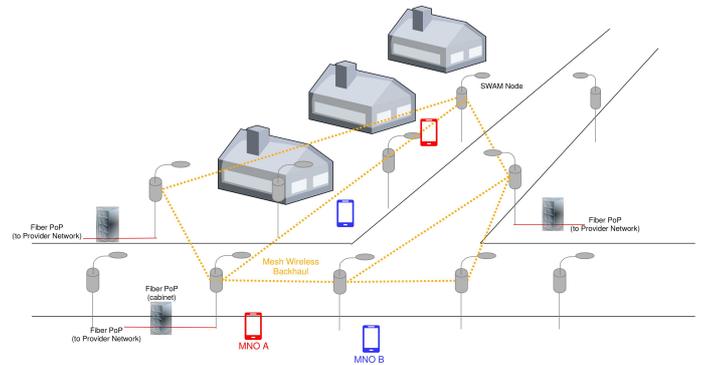}
	\caption{Dense outdoor small cell deployment with wireless access and backhaul capabilities. Clients from different MNOs connect to the network.}
	\label{fig:swam_scenario}
\end{figure}

Regarding sharing of Small Cell infrastructure, we posit that the multi-tenancy paradigm currently used by cloud providers, where a common compute infrastructure is dynamically allocated to serve different tenants, should be applied to future outdoor Small Cell deployments in order to serve multiple Mobile Network Operators (MNOs). Indeed, such capability is fully aligned with the 5G vision of \emph{slicing} \cite{5G_slicing}. Figure \ref{fig:swam_scenario} depicts an example urban scenario including a multi-tenant dense Small Cell deployment using wireless backhauling.

In this paper we present \textit{SWAM}, a system that builds on commodity Wi-Fi routers, to provide a wireless access infrastructure supporting multi-tenancy, mobility and integrated access and backhaul capabilities. SWAM can be used by an infrastructure provider to provision on-demand connectivity for different MNOs, i.e. \emph{tenants}, as depicted in Figure \ref{fig:swam_scenario}. 

Our main contributions in this paper are:
\begin{itemize}
\item The SWAM datapath and control plane, which allow to instantiate a set of per-tenant virtual access points, and backhaul them wirelessly until the tenant home network. SWAM supports advanced control plane features such as dynamic per-tenant gateways, and client mobility.
\item A SWAM prototype based on commercial wireless cards and an embedded single board computer running Linux. Our prototype is based on realistic hardware and could be deployed in the field with a proper outdoor enclosure.
\item An experimental evaluation using an indoor office testbed composed of five SWAM nodes, which faithfully represents a  realistic SWAM deployment.
\end{itemize}

%The main contribution of this paper is the design and experimental validation of the SWAM datapath and control plane, which allow to collect the traffic from the virtual access points, and deliver it to each tenant's Home Network. %The work in SWAM is inspired by the networking solutions used in current data-center networks, specially OpenStack Neutron \cite{openstack-neutron}, which however are significantly modified to fit the requirements of the embedded devices used in SWAM. 

The paper is organized as follows. Section \ref{sec:system_design} describes the SWAM datapath and its associated control plane. Section \ref{sec:perf_eval} describes our implementation of SWAM and reports on the performance achieved in our custom office testbed. Section \ref{sec:related_work} describes related works. Finally, Section \ref{sec:conclusions} summarizes and concludes the paper.

\section{System Design}
\label{sec:system_design}

In a typical implementation, a SWAM node may be equipped with three or four wireless Network Interface Cards (NICs) and one Ethernet interface in case it connects to the wired network. One or two wireless NICs are used to provide access traffic at the 2.4 GHz and/or 5 GHz band, while the other two wireless NICs are used to provide wireless backhaul functionality in the 5 GHz band. Interference between the wireless NICs is mitigated through the use of directive antennas in the backhaul, whilst omni-directional antennas are used in the access.

%\begin{figure}
%	\centering		
%        \includegraphics[width=0.5\textwidth]{figs/swam_hw.eps}
%	\caption{Example HW architecture of a SWAM node. An embedded board, equipped with mini-PCIe interfaces connecting several wireless NICs. Directive antennas are used to isolate the 5 GHz interfaces from each other.}
%	\label{fig:swam_hw}
%\end{figure}

\begin{figure}
	\centering		
        \includegraphics[width=0.5\textwidth]{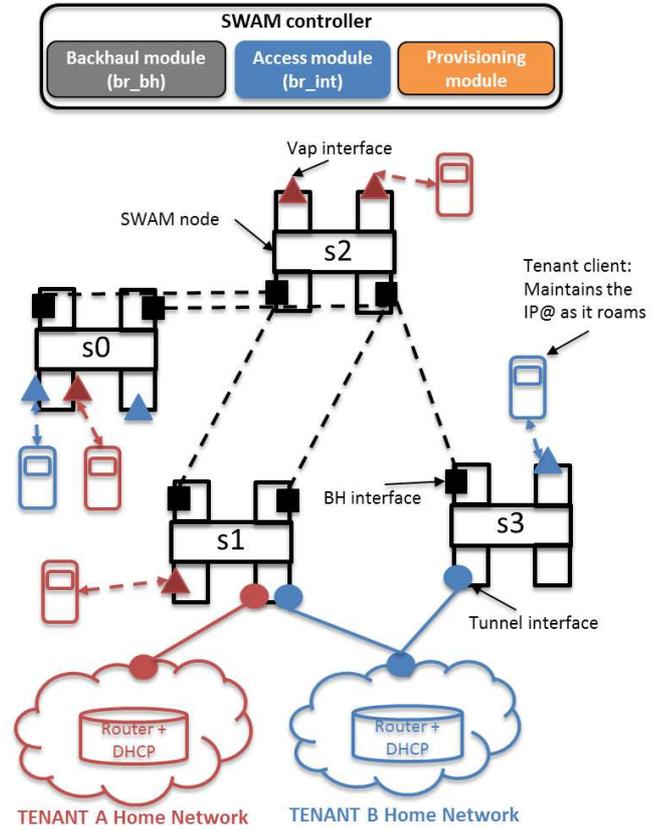}
	\caption{Sample SWAM deployment. A set of SWAM nodes instantiate per-tenant \emph{vaps}, depicted with a colored triangle, per-tenant \emph{tunnel\_if} interfaces, depicted with a colored circle, and common backhaul interfaces, depicted with a dark square. Per-tenant client devices connect to their corresponding \emph{vap} and have their traffic forwarded until the tenant's Home Network, which also provides IP address allocation. The SWAM controller manages the access and transport functions in the SWAM nodes. Figure best seen in color.}
	\label{fig:swam_deployment}
\end{figure}

SWAM accommodates multi-tenancy through the instantiation of virtual interfaces on top of the wireless access NICs, where each virtual interface represents a virtual access point (\emph{vap}), serving traffic for a particular tenant. Backhaul interfaces of SWAM nodes connect to each other forming a wireless mesh network. In SWAM, control and management of the access and backhaul resources are accomplished using the SWAM controller illustrated in Figure \ref{fig:swam_deployment}, which performs three main tasks: i) Steering traffic through the wireless backhaul (backhaul module), ii) multiplexing access traffic into the wireless backhaul (access module), and iii) dynamically provision the required virtual interfaces for each tenant (provisioning module).

Figure \ref{fig:swam_deployment} depicts an exemplar SWAM deployment where multiple tenants instantiate \emph{vaps} across the access interfaces of the SWAM nodes. The different \emph{vaps} are wirelessly backhauled to one or more nodes connected to the wired network. These nodes are hereafter referred to as SWAM gateways, and contain tunnel interfaces towards the Home Network of each tenant. Connectivity services, including IP address allocation, are hosted in each tenant's respective HN. Hence, once a customer connects through the \emph{vaps} of a given tenant, it will receive an IP address from the tenant's HN, which will be maintained while the tenant roams across the network (c.f. Figure \ref{fig:swam_deployment}).

\subsection{SWAM datapath}
\label{subsec:control_plane}

\begin{figure*}[ht]
	\centering		
        \includegraphics[width=0.8\textwidth]{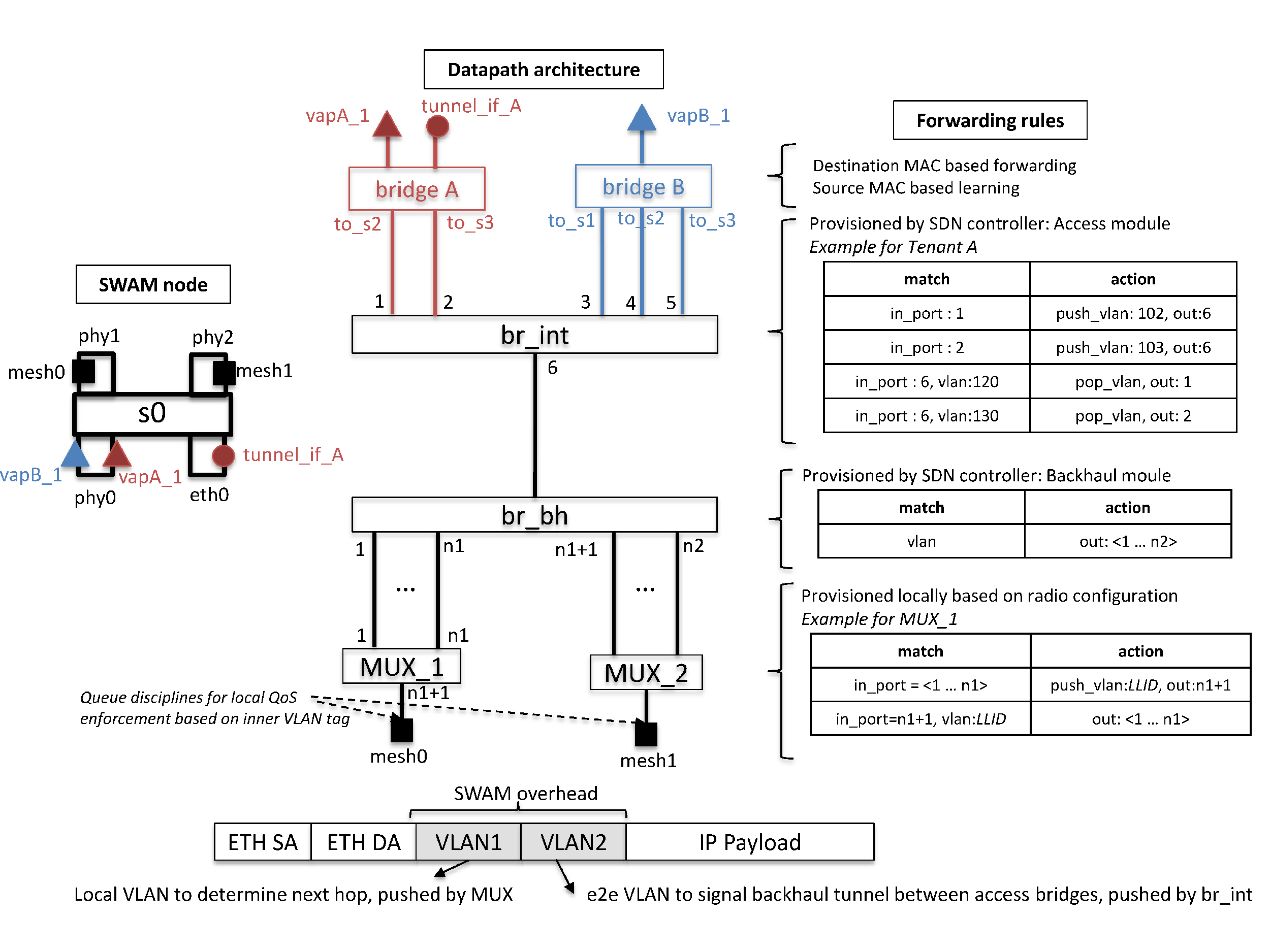}
	\caption{Example SWAM datapath deployed in SWAM node $s0$. Two tenants, $A$ and $B$, have \emph{vap} and tunnel interfaces in this node depicted with blue and red colors. Tenant $A$ has additional presence in SWAM nodes $s2$ and $s3$, and tenant $B$ in SWAM nodes $s1$, $s2$ and $s3$. The figure highlights the rules installed in the integration bridge that maps the per-tenant packets into backhaul tunnels. For simplicity tunnels IDs are encoded as $tij$, where $t$ represents the tenant and $i,j$ the origin and destination SWAM nodes. Figure best seen in color.}
	\label{fig:swam_datapath}
\end{figure*}

Figure \ref{fig:swam_datapath} depicts the datapath in charge of packet processing in a SWAM node. On the left hand side of Figure \ref{fig:swam_datapath} we can see an example of a SWAM node with three physical wireless interfaces and one Ethernet interface. One wireless interface is used to serve access traffic and instantiates two $vap$ interfaces for tenant A and B, whereas the other two wireless interfaces are used for wireless backhaul and instantiate two $mesh$ interfaces. The Ethernet interface connects to the wired network and instantiates a tunnel interface\footnote{Not all SWAM nodes have a connected wired interface, only SWAM gateways.}. 

The goal of the SWAM datapath, depicted in the middle of Figure \ref{fig:swam_datapath}, is to process packets coming from the tenants' customers ($vap$ interfaces) and deliver them to the appropriate SWAM gateways through the wireless backhaul ($mesh$ interfaces). A three level hierarchy of software switches is used for this purpose: i) Per-tenant access bridges, ii) the integration bridge (\emph{br\_int}), and iii) the backhaul bridge (\emph{br\_bh}).

The core idea behind the SWAM datapath is a logical separation between the access and the backhaul. The job of the wireless backhaul is to forward packets along a set of end-to-end tunnels, whereas the job of the access side is to match traffic coming from the tenants' \emph{vaps} to the appropriate backhaul tunnels. In SWAM, a backhaul tunnel is defined using a VLAN tag, and provides a unidirectional connection between two interfaces of a per-tenant access bridge. Figure \ref{fig:swam_datapath} depicts a sample SWAM datapath in node $s0$, where tenants A and B have instantiated \emph{vaps}, along with their corresponding access bridges ($br_{A}$ and $br_{B}$). Tenant A has also instantiated presence in SWAM nodes $s2$ and $s3$. Therefore, $br_{A}$ has two backhaul facing ports, representing respectively tunnels $s_{A:0\rightarrow 1}$ and $s_{A:0\rightarrow 2}$ for tenant A, which as depicted in Figure \ref{fig:swam_datapath} map to VLANs $102$ and $103$ in the integration bridge. Notice that backhaul tunnels are unidirectional, hence the reverse tunnels for tenant A, i.e. $s_{A:1\rightarrow 0}$ and $s_{A:2\rightarrow 0}$ map to different VLANs ($120$ and $130$ in Figure \ref{fig:swam_datapath}). 

We can now describe the detailed processing of each software bridge.

\subsubsection{Per-tenant bridges}
In a SWAM node there is one access bridge, $br_{i}$, for each tenant that has presence in that node. The per-tenant bridge manages the following interfaces: i) all the \emph{vaps} for this tenant, ii) any \emph{tunnel\_if} for this tenant, and iii) a set of virtual interfaces connecting to the other SWAM nodes where this tenant has presence. The per-tenant bridges behave as traditional MAC learning bridges, hence they dynamically learn the location of other tenant clients in the network.

\subsubsection{Integration bridge}
The job of the integration bridge, $br\_int$, is to map traffic from the tenants' access and home networks to the wireless backhaul tunnels. In particular, $br\_int$ maintains a mapping between each backhaul facing port of the per-tenant access bridges and their corresponding backhaul tunnel. To implement this binding, $br\_int$ pushes and pops the corresponding tunnel VLANs according to the interface of the incoming traffic. The right part of Figure \ref{fig:swam_datapath} depicts the portion of the forwarding table in $br\_int$ affecting Tenant A.

The forwarding table in the integration bridge is populated by the access module in the SWAM controller using OpenFlow \cite{openflow}. 
The provisioned rules are expected to be fairly static in practice, since they only need to be provisioned when a tenant requests to instantiate presence in a new SWAM node. 
%The SWAM provisioning interface is described in Section \ref{subsec:swam_mgmt}. 
In Section \ref{subsec:swam_control} we will discuss the need of additional rules in $br\_int$ to prevent network loops and enable load balancing.

\subsubsection{Backhaul bridge}
The job of the backhaul bridge, $br\_bh$, is to switch backhaul tunnels based on their VLAN identifier, as depicted in the $br\_bh$ forwarding table in the right part of Figure \ref{fig:swam_datapath}. There is a subtlety to be considered though. SDN software switches, as $br\_bh$, have been designed to control point-to-point (p2p) physical interfaces, whereby the forwarding table determines the next-hop for a packet by deciding the output interface. However, the  wireless interfaces used for backhauling purposes in SWAM are point-to-multipoint (p2mp). Hence, we cannot attach a wireless interface directly to $br\_bh$, since $br\_bh$ would not be able to specify what the next hop should be. Instead, we adopt the SDN wireless architecture proposed in \cite{sesame_icc}, whereby each potential destination reachable through a p2mp wireless interface is represented using a virtual Ethernet interface, which is attached to the SDN software switch ($br\_bh$). In this way, $br\_bh$ can decide the next hop for each backhaul tunnel by forwarding through the appropriate virtual interface. An additional software switch is required for each physical wireless interface, $\textnormal{\textit{MUX}}_{i}$, which multiplexes the traffic coming from $br\_bh$ into each physical wireless interface by adding an outer VLAN tag. This outer VLAN tag is determined locally in each SWAM node based on the Local Link ID (LLID) parameter used in the wireless MAC layer (IEEE 802.11s), and is used by the driver of the wireless interface to determine the next hop. The interested reader is referred to \cite{sesame_icc} for a detailed description of this architecture.

The forwarding table of $br\_bh$ is populated by the backhaul module in the SWAM controller using the OpenFlow protocol (c.f. Figure \ref{fig:swam_deployment}). The actual path followed by each tunnel along the wireless backhaul can be modified by the SWAM controller, simply by acting on the $br\_bh$ bridges of the affected SWAM nodes, without the access bridges or the integration bridge being affected. The definition of algorithms to control forwarding of the backhaul tunnels is out of the scope of this paper, but the interested reader is referred to \cite{eucnc_16} for a representative example. In addition to traffic engineering across the wireless backahul, SWAM nodes also support local per-tunnel QoS enforcement by applying queue disciplines to the output wireless backhaul interfaces. Queue disciplines are applied on the inner VLAN tag representing the end-to-end tunnel (c.f. Figure \ref{fig:swam_datapath}).

Finally, it is worth noticing that, in order to achieve the required multi-tenancy, SWAM imposes a minimal overhead: only two stacked VLAN tags as depicted in Figure \ref{fig:swam_datapath}.

\subsection{Scalability of the SWAM datapath}

The factors limiting the scalability of SWAM, are the number of concurrent tenants, the number of backhaul tunnels, and the number of rules maintained in the SWAM datapath.

The maximum number of concurrent \emph{vaps} over a physical wireless NIC is limited by the overhead introduced by each \emph{vap}. Typically, a \emph{vap} transmits a Beacon frame every 100~ms with a slow Modulation and Coding Scheme (MCS). Enterprise Wi-Fi vendors currently recommend a maximum number of $5$ \emph{vaps} per wireless NIC \cite{max_vaps}, although the number could be increased using a higher MCS for Beacon frames. We consider an appropriate number of concurrent tenants in SWAM to be between $5$ and $10$.

Using VLAN IDs as tunnel identifiers, SWAM can have a maximum of $4096$ unidirectional backhaul tunnels. Considering $T$ the number of tenants in the network, and $N$ the number of SWAM nodes, a maximum of $2TN(N-1)$ unidirectional backhaul tunnels are required if all tenant have presence in all SWAM nodes. As a matter of example, if we assume respectively $T=10$ and $T=5$ concurrent tenants, then $N=14$ and $N=20$ SWAM nodes can be accommodated in the same SWAM network. We consider this network size reasonable, given the application of SWAM in outdoor Small Cell deployments. However, bigger network sizes can still be accommodated by: i) using a more complex backhaul tunnel identifier (e.g. double tagging only for the tunnel), or ii) partitioning the network in independent control areas as proposed in \cite{5gxhaul_slicing}, each with a dedicated SWAM controller.

Finally, notice that using software switches a significant number of rules can be maintained in the SWAM datapath. However, operating with a small number of rules is still beneficial because SWAM nodes use embedded platforms with limited memory, and wireless interfaces approaching gigabit speeds may stretch the performance of software switches. Under the previous assumptions, the total number of rules to be maintained in the SWAM datapath are: i) ${O(2TN(N-1))}$ rules in the $br\_bh$, ii) ${O(2T(N-1))}$ rules in the $br\_int$, and iii) ${O(Clients)}$ entries in the MAC tables of each of the $T$ per-tenant bridges, where $Clients$ is the number of client devices active in the network for a given tenant. The critical factor is thus $Clients$, which may be large in the gateway nodes that need to maintain the point of attachment for all clients in the network. Hence, one may have to equip gateway nodes with additional memory or balance the role of gateway for different tenants, as explained in Section \ref{subsec:swam_control}.

\subsection{SWAM control plane}
\label{subsec:swam_control}

%The SWAM control plane populates the datapaths described in the previous section. SWAM uses a combination of distributed and centralized control plane. On the one hand, the per-tenant bridges implement data-plane learning in a distributed fashion. On the other hand, the integration and backhaul bridges are controlled, using the OpenFlow protocol \cite{openflow}, respectively by an access and backhaul modules in the logically centralized SWAM controller, as depicted in Figure \ref{fig:swam_deployment}. 

%In order to forward packets, \emph{br\_int} is pre-provisioned with a set of static rules that bind the input ports coming from the per-tenant bridges, into a specific backhaul tunnel identifier, i.e. VLANs in our implementation. Figure \ref{fig:swam_datapath} provides a detailed description of the incoming and outgoing rules installed in \emph{br\_int}. 

%Backhaul tunnels between any pair of SWAM nodes are pre-provisioned in \emph{br\_bh} by the backhaul module in the SWAM controller. However, the actual path followed by these tunnels can be dynamically updated depending on the conditions of the wireless backhaul. An in-depth description of the algorithms used in the SWAM controller to program the backhaul tunnels is out of the scope of this paper, however the interested reader is referred to \cite{eucnc_16} for a relevant example.

As a consequence of the SWAM datapath introduced in the previous section, a given tenant enjoys a layer two abstraction whereby its access bridges are connected to each other forming a mesh, as illustrated in the left part of Figure \ref{fig:single_multi_gw}. As we can see the resulting overlay topology contains loops, which is a problem given that the access bridges are regular MAC learning bridges. SWAM contains control plane mechanisms that provide three main features: i) avoid loops in the resulting per-tenant overlays, ii) enable the possibility of using multiple gateways for a given tenant to balance load across the wireless backhaul, and iii) support client mobility, whereby when a client device hands over from one tenant \emph{vap} to another the backhaul tunnels are re-configured appropriately.

\subsubsection{Loop avoidance and multiple gateways support}
\label{subsec:mult_gws}

\begin{figure}
	\centering		
        \includegraphics[width=0.5\textwidth]{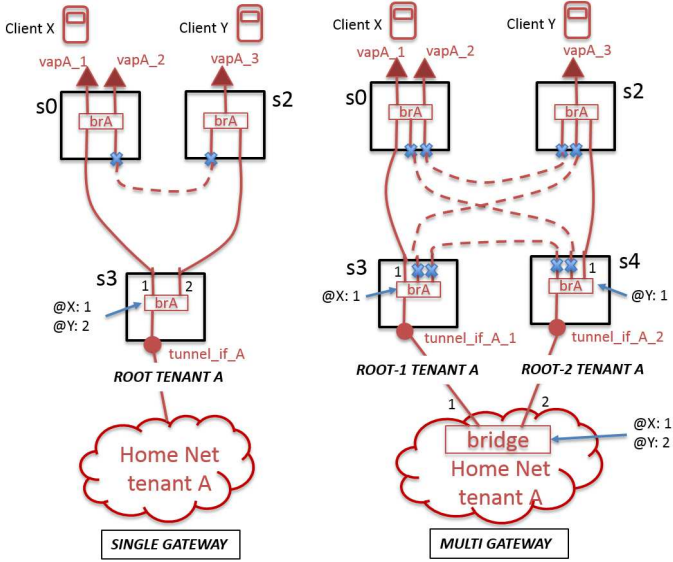}
	\caption{Left. Resulting overlay from the perspective of a single tenant. The SWAM controller blocks ports to avoid loops. Right. Situation when considering multiple gateways (root nodes). The bridge at the tenant's home network binds the MAC address from the client devices to the port connected to the corresponding gateway.}
	\label{fig:single_multi_gw}
\end{figure}

To avoid loops in the resulting tenant overlay the access module in the SWAM controller implements a traditional Spanning Tree algorithm to each tenant overlay. To apply Spanning Tree one SWAM node for each tenant needs to be appointed as the \emph{root} node. In SWAM the tenant root node is a SWAM gateway node having a tunnel interface to the tenant home network (c.f. Figure~\ref{fig:swam_deployment}). Consequently, in the access bridges in the SWAM nodes for tenant $t$, we need to block all the ports $p_{<t:i\rightarrow j>}$ linking to a backhaul tunnel other than the tunnel to the tenant root, i.e. $p_{<t:i\rightarrow r_{t}>}$, where $r_{t} \in S$ is the SWAM node acting as root for tenant $t$, and $S=\{s_{0}, ... ,s_{|S|-1}\}$ is the set of all SWAM nodes. In order to block the ports in the SWAM datapath, the access module in the SWAM controller pushes a high priority drop rule into the integration bridge linked to the appropriate access port (c.f. Figure \ref{fig:swam_datapath}). Since all SWAM nodes have a direct tunnel to the root node, the resulting per-tenant overlay is a hub-and-spoke topology as depicted in the left part of Figure~\ref{fig:single_multi_gw}.

Introducing a hub-and-spoke architecture renders direct communications between devices connected to different \emph{vaps} inefficient, since they have to go through the root node, consuming expensive wireless backhaul resources. However, we argue that this is not a relevant problem in practice, since in many wireless hotspots direct communication between devices is forbidden anyway for security reasons, and in practice all traffic is directed to the tenant home network to connect to the Internet. However, using a single root node per tenant forces all tenant traffic to go through the same SWAM gateway, which can introduce congestion in the wireless backhaul. 

To enable load balancing, SWAM supports the allocation of multiple concurrent gateways for a given tenant. Multiple gateways are enabled simply by configuring the tenant access bridge in different non-root SWAM nodes to point to a different SWAM gateway using the appropriate drop rules in $br\_int$. In this way, the resulting tenant overlay is partitioned in multiple sub-trees routed in the respective root nodes. In order to enable full network communication, the resulting overlay sub-trees are bridged in the tenant home network, using a regular bridge (not controlled by the SWAM controller). The right part of Figure~\ref{fig:single_multi_gw} illustrates the multi-gateway configuration.

%However, the loop-free architecture also hinders the ability of having multiple separate gateways for a given tenant, which is highly beneficial to spread the load across the wireless backhaul, thus alleviating congestion points. SWAM supports multiple gateways for a given tenant by electing multiple root nodes, and assigning each SWAM node where the tenant has presence, to one of the root nodes. In this case though, an additional bridge is required in the tenant's HN that directs downstream traffic to the proper SWAM gateway, as depicted in the right part of Figure \ref{fig:single_multi_gw}. 

The access module in the SWAM controller may decide to vary the number of root nodes for each tenant dynamically according to traffic carried by the wireless backhaul. However, there is a caveat to be considered to minimize the impact of updating the root node on the ongoing sessions. Given the resulting hub-and-spoke topology perceived by each tenant in SWAM, the point of attachment of each tenant client device is maintained in the tenant access bridge of the root node, i.e. the MAC address of each client, and in the tenant's home network bridge if multiple gateways are used for that tenant (c.f. $@X$ and $@Y$ in Figure \ref{fig:single_multi_gw}). Therefore, if the access module in the SWAM controller updates the root node in a particular SWAM node, the access bridge hosting the client \emph{vap} and the bridge in the tenant home network need to be updated. In order to update the MAC entries in these bridges, SWAM nodes host a control plane agent that spoofs an ARP request on behalf of the connected clients. This agent is triggered by the access module in the SWAM controller. Algorithm \ref{alg:update_tenant_root} describes the procedure employed by the SWAM controller to update the root node for a tenant. In Section \ref{sec:perf_eval} we will evaluate the performance of this scheme. 

%The number of gateways used by a particular tenant can be varied dynamically. Dynamically allocating a gateway node only requires to modify the drop rules in the integration bridge of the affected SWAM nodes. In Section \ref{sec:perf_eval} we demonstrate the benefits of dynamically allocating multiple gateways in a simplified network topology. We leave for future work the design of more complex algorithms to accomplish this task.

\begin{algorithm}
\DontPrintSemicolon % Some LaTeX compilers require you to use \dontprintsemicolon instead
// Variables\; 
$S=\{s_{0}, ... ,s_{|S|-1}\}$ //Set of SWAM nodes\; 
$T=\{t_{0}, ... ,t_{|T|-1}\}$ //Set of tenants in the network\; 
$r\_old_{t} \in S$ //Current root node for tenant $t$\; 
$r\_new_{t} \in S$ //Target root node 
for tenant $t$\; 
$p_{<t:i\rightarrow j>}$ //Port in $br\_int$ linked to the $s_{t:i\rightarrow j}$ tunnel\;
$br_{t}.vaps$ //\emph{vaps} in access bridge for tenant $t$\;
\;
//Update drop rules in $br\_int$ for new root node\;
$p_{<t:i\rightarrow r\_old_{t}>} \gets drop$\;
$p_{<t:i\rightarrow r\_new_{t}>} \gets enabled$\;
\;
//Spoof ARP Request for attached clients\;
$Clients \gets acc\_br_{t}.vaps.maclist$\;
\For{$mac \in Clients$} {
  $arp\_req.sa \gets mac$\;
  $arp\_req.da \gets broadcast$\;
  $trigger \ ARP \ Request$\;
}
\caption{update\_tenant\_root()}
\label{alg:update_tenant_root}
\end{algorithm}

\subsubsection{Mobility support}

A client device attached to a \emph{vap} of a given tenant should maintain connectivity while roaming through the network. There are two mechanisms involved in maintaining connectivity. First, the process of executing a handover between \emph{vaps}, which is a standard feature supported by the Wi-Fi devices used in SWAM. Second, a client handover triggers an update of the path followed by the packets addressed to the client through the wireless backhaul. %To understand how SWAM is to understand which entity maintains the location of each client in the network

Similarly to the tenant root re-allocation mechanism described in the previous section, in order to maintain connectivity upon a client handover, the MAC lists in the access bridge in tenant root node and in the home network bridge, need to be updated in order to point to the tunnel connecting to the SWAM node where the client is currently attached. Handovers in SWAM are \emph{break before make}, since the SWAM controller has no control over the client devices in order to instruct them when to execute the handover. Therefore, it is critical to properly update the MAC lists in the affected bridges as soon as a new point of attachment (\emph{vap}) for a client device is detected. There are two mechanisms in SWAM to accomplish this. First, the control agent in the SWAM nodes is notified by the \emph{vaps} when a new client attaches, and generates a spoofed ARP Requests as described in Section \ref{subsec:mult_gws}. Second, we have observed that clients upon re-association, spontaneously generate a broadcast layer-two packet for link discovery purposes (LLC-xID). This packet is forwarded to the root node through the newly attached tunnel, hence automatically updating the MAC tables appropriately. In Section \ref{sec:perf_eval} we will evaluate impact of SWAM handovers on ongoing traffic.

\section{Experimental evaluation}
\label{sec:perf_eval}
\begin{figure}
	\centering		
        \includegraphics[width=0.35\textwidth]{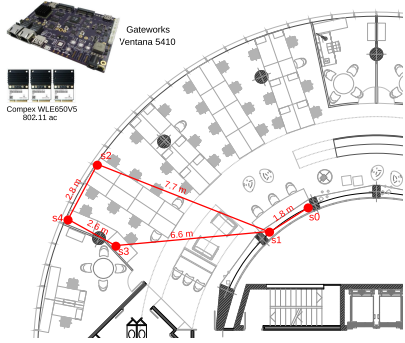}
	\caption{Physical position of the 5 nodes within our office along with the wireless backhaul links and distances between the connected nodes.}
	\label{fig:swam_indoor_deployment}
\end{figure}

\begin{figure}
	\centering		
        \includegraphics[width=0.5\textwidth]{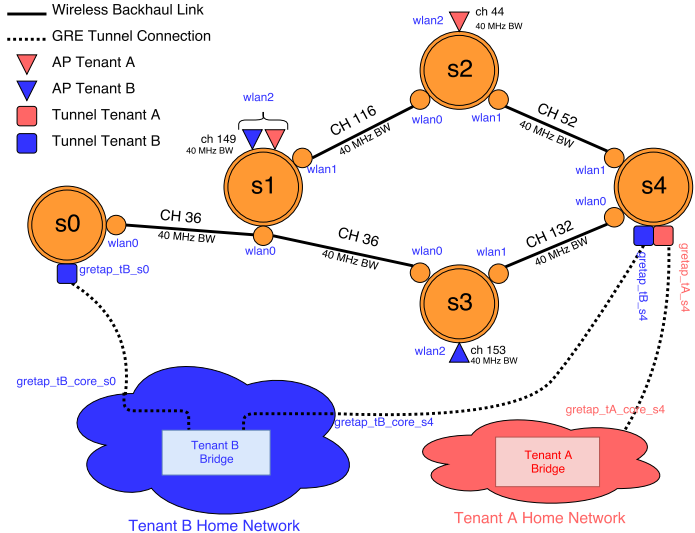}
	\caption{Logical topology used to demonstrate SWAM capabilities. Tenants \emph{A} and \emph{B} instantiate \emph{vaps} for their customer over the SWAM infrastructure, which connects clients to each tenant's home network.}
	\label{fig:swam_evaluation}
\end{figure}

The SWAM architecture is implemented using OpenvSwitch switches. 
Node $s2$ is connected to a remote server were the SWAM controller, which is implemented using OpenDayLight, runs.
In order to validate the SWAM architecture, we perform a series of experiments that showcase its functionality and performance during certain network events and under varying network conditions.
For these evaluations, we set up a physical indoor testbed composed of five SWAM nodes built with the components shown in Fig.~\ref{fig:swam_indoor_deployment}.
Each node is equipped with one IEEE 802.11ac wireless NIC to provide Wi-Fi connectivity to clients, and one or two IEEE 802.11ac NICs to establish wireless backhaul links. 
The layout of the five nodes within our office environment is depicted in Fig.~\ref{fig:swam_indoor_deployment}. Due to space constraints all the antennas used are omni-directional, unitary gain dipoles.
Using MAC layer filtering, we set up a logical \emph{kite} topology and instantiate tenant A and tenant B, with their respective \emph{vaps} and tunnel interfaces connecting to their home networks, as depicted in Fig.~\ref{fig:swam_evaluation}.

In this initial configuration, nodes $s0$ to $s4$ announce SSIDs for the two tenants on dedicated \emph{vaps}.
In particular, tenant A instantiates \emph{vaps} in $s1$ and $s2$, whereas tenant B instantiates \emph{vaps} in nodes $s1$ and $s3$.
The wireless channels for the access and backhaul networks have been chosen to maximize spatial reuse in the 5~GHz band as depicted in Fig.~\ref{fig:swam_evaluation}. The channels are configured to use 40~MHz bandwidth to provide higher link capacity.

For tenant A, $s4$ acts as gateway, while tenant B has two possible gateways: $s0$ and $s4$.
In the initial configuration, the backhaul tunnels are configured in such a way that the default gateway to reach both tenants' home networks is $s4$, leaving $s0$ as alternative gateway for tenant B. 

We connect three laptops (clients) equipped with IEEE 802.11b/g/n Wi-Fi dongles to different per-tenant \emph{vaps} of the network: on one hand, the device $\textnormal{\textit{STA}}_{A_1}$ is attached to tenant's A network announced at $s1$.
On the other hand, the devices $\textnormal{\textit{STA}}_{B_1}$ and $\textnormal{\textit{STA}}_{B_2}$ are attached to the \emph{vaps} of tenant B on nodes $s1$ and $s3$, respectively.

The default upstream path followed for data flows from $\textnormal{\textit{STA}}_{A_1}$ towards its home network goes from $s1$ over $s2$ to $s4$ (short: $s1 \! \! \rightarrow \! \! s2 \! \! \rightarrow \! \! s4$), the \textit{upper branch} of the topology. 
For simplicity, in all of the experiments the downstream path is chosen to be symmetric to the upstream path.
Thus, the backhaul tunnel $s4 \! \! \rightarrow \! \! s2 \! \! \rightarrow \! \! s1$ is used for any downstream traffic coming from tenant A's home network directed towards $\textnormal{\textit{STA}}_{A_1}$.
Similarly, for $\textnormal{\textit{STA}}_{B_1}$  a symmetric (up- and downstream) data path is instantiated over the \textit{lower branch} $s1$-$s3$-$s4$, whereas for $\textnormal{\textit{STA}}_{B_2}$ the symmetric data path goes over $s3$-$s4$.

%The SWAM controller responsible for managing, monitoring and maintaining the network runs as a virtual machine in a dedicated cloud and is connected to $s2$ via a dedicated VPN, from where all other nodes of the network can be reached over dedicated control paths in the wireless backhaul. 
%In this set up, $s2$ acts as a gateway for the other nodes that can reach $s2$ over a dedicate control path in the wireless backhaul. 
Parting from this initial configuration, three consecutive experiments are performed to evaluate how the SWAM architecture enables the following key features: i) SDN-based fast route repair, ii) load balancing mechanisms and iii) client handover processes.
Each experiment is repeated at least 40 times under the same testbed conditions.
The results presented in the following subsections are averages across these repetitions.

\subsection{Access and backhaul isolation}
In the first experiment, we demonstrate the advantages of the separation of the access and backhaul embedded in the SWAM architecture. In particular we show how SWAM reacts when a backhaul link carrying an end-to-end tunnel between two per-tenant access bridges breaks.
To enable fast recovery, SWAM proactively installs backup paths and reallocates the end-to-end backhaul tunnel.
The interested reader is referred to \cite{flrr} for a detailed description of this fast recovery scheme. 

\begin{figure} %This figure will contain the initial topology and the link-breaking event
	\centering		
        \includegraphics[width=0.5\textwidth]{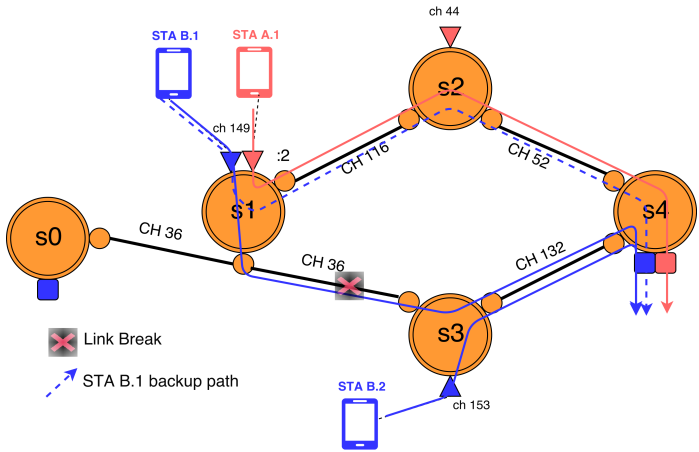}
	\caption{End-to-end upstream data flows for the 3 STAs in the initial experiment configuration, and flow reallocation due to link-breaking event.}
	\label{fig:SWAM_initial_situation}
\end{figure}

At the beginning of this experiment, an iperf client is launched on each of the three clients, generating an UDP stream with a fixed data rate of 32~Mbps towards an iperf server located in the corresponding tenant's home network. 
As stated above, $\textnormal{\textit{STA}}_{A_1}$ uses the upper branch of the topology to reach tenant A's Home Network, while $\textnormal{\textit{STA}}_{B_1}$ and $\textnormal{\textit{STA}}_{B_2}$ use the lower branch, as depicted in Fig.~\ref{fig:SWAM_initial_situation}.
% \begin{figure} %This figure will contain the initial topology and the link-breaking event
% 	\centering		
%         \includegraphics[width=0.5\textwidth]{NewFigures/SWAM_initial_situation.pdf}
% 	\caption{End-to-end upstream data flows for the 3 STAs in the initial experiment configuration. \textcolor{red}{Can we simplify this, and show the various experiments on a single diagram?, or in two.}}
% 	\label{fig:SWAM_initial_situation}
% \end{figure}

\begin{figure}
	\centering		
        \includegraphics[width=0.5\textwidth]{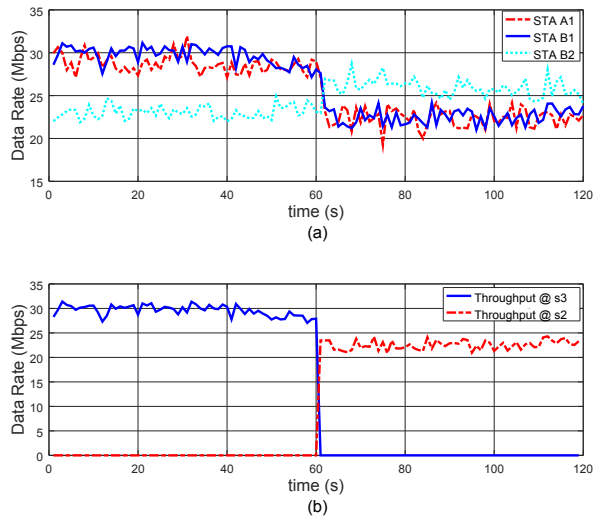}
	\caption{(a) Data Rate evolution during a backhaul link breaking event and data flow reallocation process, (b) isolated throughput of $\textnormal{\textit{STA}}_{B_1}$ measured at $s2$ and $s3$.}
	\label{fig:Iperf3clients}
\end{figure}

The iperf data rate limits have been chosen in such a way that at least two flows are necessary to saturate any of the wireless backhaul links, whereas a single flow can be carried without any issues.
At the beginning of the experiment, around 30~Mbps of end-to-end throughput for the flows originating in $s1$ ($\textnormal{\textit{STA}}_{A_1}$, $\textnormal{\textit{STA}}_{B_1}$) and around 23~Mbps for the flow originating in $s3$ ($\textnormal{\textit{STA}}_{B_2}$) are noted, as shown in Fig.\ref{fig:Iperf3clients}. 
The lower throughput observed for $\textnormal{\textit{STA}}_{B_2}$ can be explained on one hand due to saturation of the backhaul link $s3$-$s4$ and on the other hand due to cross-channel interference at $s3$.
We observe this type of interference, as there are 3 NICs on $s3$ (backhaul and access) that are active at the same time.
It is also important to note that the experiments take place in a typical office environment with other ongoing, external wireless communications.
The wireless environment is affected by these conditions and the access links behave differently for every client.
In particular, we observe that the access link $\textnormal{\textit{STA}}_{B_2}$ is not interference free. 
Another consequence of this dynamic environment, apart from reduced throughputs, are the throughput fluctuations that are measured during the entirety of the experiments. % can be observed in Fig.\ref{fig:Iperf3clients} (a).
This type of interference is not expected to occur in outdoor deployments, when SWAM nodes use directive antennas in the backhaul, allowing for more robust and steadier links.

The link break to which SWAM has to react happens 60~seconds after initiating the experiment.
To trigger the link break, the wireless backhauling interface on $s3$ that connects with $s1$ is forcefully shut down, interrupting the upstream data flow generated by $\textnormal{\textit{STA}}_{B_1}$. 
%A Fast and Local Re-Routing (FLRR) agent software \cite{flrr} running on $s1$ detects the interruption of the connection in the backhaul link.
SWAM immediately reacts by reallocating $\textnormal{\textit{STA}}_{B_1}$'s traffic to the only possible backup path towards $s4$ over the upper branch, as represented with a dashed line in Fig.~\ref{fig:SWAM_initial_situation}.
% \begin{figure} %Get rid of this figure, The new reference will be ...
% 	\centering		
%         \includegraphics[width=0.5\textwidth]{NewFigures/SWAM_reallocation.pdf}
% 	\caption{Backhaul flow reallocation after the link breaking event.}
% 	\label{fig:SWAM_reallocation}
% \end{figure}

In average, we measure a path reallocation time of around 246~ms for $\textnormal{\textit{STA}}_{B_1}$'s flow, during which packets are dropped.
This connectivity timeout leads to a temporary decrease of its throughput. 
This barely noticeable drop can be seen in Fig.~\ref{fig:Iperf3clients}(b) that shows the isolated throughput of $\textnormal{\textit{STA}}_{B_1}$ measured at nodes $s3$ and $s2$ before and after the reallocation process during a selected experiment. 
At the 60~second mark it can be seen how the end-to-end flow stops flowing through $s3$ and how it is reallocated on to $s2$, revealing a seamless reallocation procedure.
While in this experiment it is shown that the quick reallocation process executed by SWAM has barely an impact on the performance of $\textnormal{\textit{STA}}_{B_1}$'s end-to-end flow, it eventually affects the overall network conditions.
On one hand, $\textnormal{\textit{STA}}_{B_2}$ experiences an increase of its end-to-end throughput, as the interference from link $s1$-$s3$ is removed.
Also, $s3$ suffers from less local radio interference as the two still active NICs no longer compete with the deactivated NIC that was used for the backhaul link $s1$-$s3$.
On the other hand, $\textnormal{\textit{STA}}_{B_1}$ and $\textnormal{\textit{STA}}_{A_1}$ now use the same backhaul links and they have to share the available capacity. 
Since the sum of the two end-to-end flows exceeds the link capacity, we observe an overall reduction of their throughput to around 23~Mbps. 
SWAM detects this network congestion, which leads to the next experiment, where the network congestion is reduced via a gateway reallocation triggered by the SWAM controller.
 
%which causes a temporary disconnection and thus a throughput drop.
%$STA_{A_1}$ and $STA_{B_1}$ data flows experience a sudden drop in performance since the backhaul module in the SDN-Controller has reallocated both the flows on the same network branch, saturating the link capacity. On the other hand $STA_{B_2}$ data rate increases when the $s3$-$s4$ link become free from $STA_{B_1}$ flow.

%, resulting in an average throughput decrease of 678~Kbps. 
%Notice that the access side of the topology remains unaware about the changes happened in the backhaul, as there are no interruptions or lost connection issues. 
%As a matter of fact, the throughput drop lies within the range of the standard data rate fluctuations of the radio channel.

\subsection{Gateway reallocation}
During the initial flow allocation, as described in the previous experiment, the overall traffic carried by the three end-to-end tunnels is fluctuating around 83~Mbps (Fig. \ref{fig:Iperf3clients}(a)). 
SWAM's reaction to the link break ensures a quick reallocation of any affected end-to-end flow.
However, the reallocation of the data flow causes the upper branch of the topology to become congested, reducing the overall carried traffic to around 73~Mbps, 10~Mbps lower than in the initial state. 
This section evaluates how, in order to alleviate this congestion, the SWAM controller decides to balance the traffic flows in the network.

\begin{figure} %
	\centering		
        \includegraphics[width=0.5\textwidth]{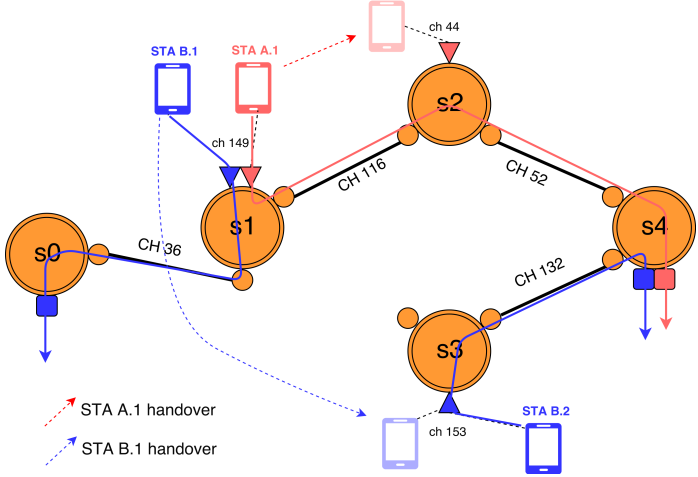}
	\caption{End-to-end flow allocation after the gateway relocation, and after the clients' handover.}
	\label{fig:SWAM_GWreallocation_mobility}
\end{figure}

Since tenant B has multiple gateways, there are several options to direct the traffic generated by tenant B's clients across the backhaul towards the home network.
The SWAM controller determines that, given the global network situation, network load balancing is possible by reallocating the root node for tenant B in node $s1$, from $s4$ to $s0$, while maintaining $s4$ as the root node for tenant B in $s3$ (affecting $\textnormal{\textit{STA}}_{B_2}$). 
In this way, $\textnormal{\textit{STA}}_{B_1}$ flow reaches the tenant B's home network through the unused backhaul link $s1\! \! \rightarrow \! \! s0$, freeing the upper branch, as depicted in Fig.\ref{fig:SWAM_GWreallocation_mobility}, and balancing the data flows across the network.
The gateway relocation process is performed as described in Algorithm 1.

%by changing the backhaul tunnels associated to the up- and downstream flows of $\textnormal{\textit{STA}}_{B_1}$, which requires the SDN-Controller to take two actions at the integration bridge level: first, the SDN-Controller modifies the OvS ruleset of the integration bridge at $s1$ via OF to make any traffic coming from tenant B's \emph{vap} use the backhaul tunnel that connects with $s0$ (by marking the packets with the appropriate backhaul tunnel VLAN).
%Second, the controller spoofs an ARP request packet and injects it into the integration bridge of $s1$.
%The ARP packet is intended to update [TODO: skip next part of sentenceM if yes, check sentence for grammar] [the tenant bridge in $s1$ and] the bridge located in the tenant home network, both self-learning switches that are not directly controlled by the SDN controller.

%The spoofed ARP request packet carries the client's MAC source address and the destination IP address of the tenant's home network. 
%The generation of the packet carrying this spoofed information is possible, since the client's MAC address and the tenant IP are recorded by the SDN-Controller during the client joining process, when the client IP address is assigned by the tenant's DHCP server. 
As the spoofed broadcast ARP request packet floods the network, it updates the port through which $\textnormal{\textit{STA}}_{B_1}$ can be reached in the MAC tables of all the MAC learning switches, eventually reaching the tenant's home network (downlink path update).
In the downstream direction, the ARP reply containing the MAC address of the tenant's home network updates, on its way back to $s1$, the per-tenant access bridges of the network, allowing to maintain the reachability between the client and the tenant's home network over the newly associated gateway and the corresponding backhaul link.

% In order to test the performance of the network in the occurrence of gateway relocation, reachability tests are performed. 
% The reachability tests consist in sending ICMP reachability packets (pings) from $\textnormal{\textit{STA}}_{B_1}$ to the tenant B home network every 5~ms. 
% During the pinging, the gateway for $\textnormal{\textit{STA}}_{B_1}$ is relocated and the Round-Trip Time (RTT) is measured. 
% The graph shown in Fig.\ref{fig:RTT_gw_reallocation} represents the average RTT measured for each ping across the experiments.
To evaluate the impact of relocating a gateway on network performance, we perform two experiments. First, in order to measure the time the network is in outage, we send a continuous stream of ICMP messages every 5~ms using the ping tool and measure the RTT. 
Second, we study how a gateway reallocation affects an ongoing TCP stream.
\begin{figure}
	\centering		
        \includegraphics[width=0.5\textwidth]{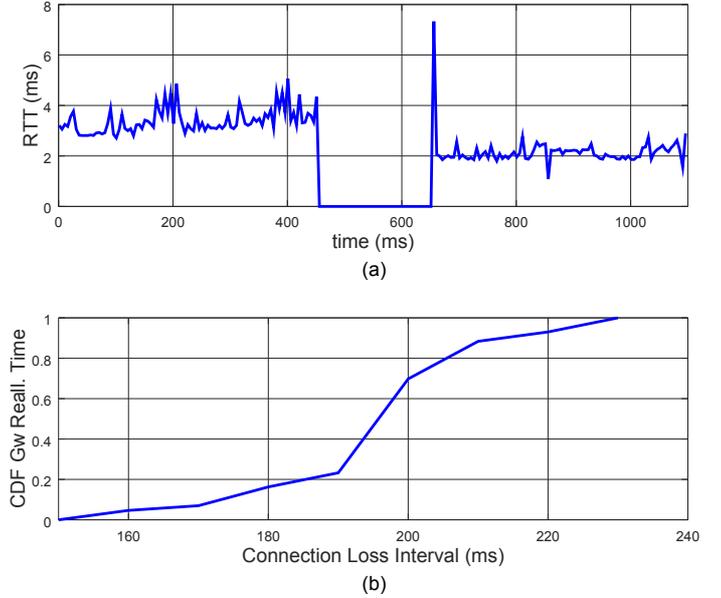}
	\caption{(a) CDF of the RTT and (b) average RTT measured when $\textnormal{\textit{STA}}_{B_1}$ gateway relocation occurs.}
	\label{fig:RTT_gw_reallocation}
    \vspace{-0.3cm}
\end{figure}

Figure \ref{fig:RTT_gw_reallocation}(a) depicts the result from one of the ICMP experiments.
At the beginning, when the gateway for $\textnormal{\textit{STA}}_{B_1}$ is $s4$, the average RTT fluctuates around 4~ms, as the end-to-end flow traverses two hops in the wireless backhaul ($s1$-$s2$-$s4$). 
After the gateway relocation, $\textnormal{\textit{STA}}_{B_1}$'s flow crosses only one backhaul link ($s1$-$s0$) before reaching the tunnel interface to tenant B's home network, and the resulting RTT decreases to 2~ms. 
However, the gateway relocation process is not immediate, as the rule update and ARP spoofing performed by the SWAM controller require a certain amount of time for processing and packet transmissions.
%The SDN-Controller has to update the rules in the integration bridges of $s1$ and $s4$ both for the upstream and downstream flow, and spoof the ARP packet to update the backlearning switches, moreover the ARP replay has to be generated by the tenant home network and received by the backlearning switches in order to update their MAC table. 
The whole relocation takes an average of 201.72~ms, the time measured between the moment in which the last valid ICMP packet reaches the tenant's home network via gateway node $s4$ and the time when the first valid ICMP packet reaches the tenant's home network through gateway node $s0$.
Figure~\ref{fig:SWAM_GWreallocation_mobility}(b) shows the empirical CDF of the connection outage times resulting from the relocation process measured in all the experiments.
Note that the relocation time depends on the performance and the delay of the control path between the nodes and controller.%\footnote{Note that running the controller on dedicated hardware and close to the network rather than on a VM and a connection via VPN would reduce the relocation time further.}. 

\begin{figure}
	\centering		
        \includegraphics[width=0.5\textwidth]{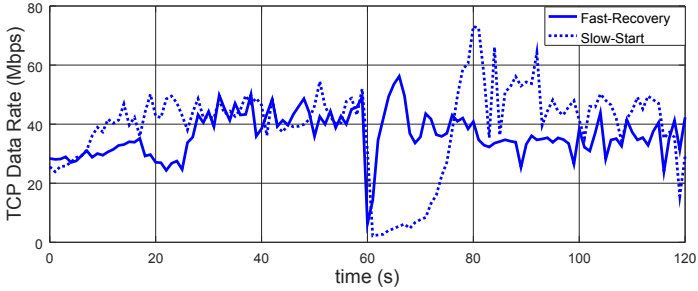}%GwReallocation_TCPupstreamResizedCaption.pdf}
	\caption{TCP throughput measured when $\textnormal{\textit{STA}}_{B_1}$ gateway relocation occurs with fast recovery (continuous plot) and slow start (dashed plot).}
	\label{fig:TCP_gw_reallocation}
\end{figure}

In our second experiment, an iperf with TCP traffic is launched towards tenant B home network, in order to determine the impact of the reallocation process on the end-to-end throughput of $\textnormal{\textit{STA}}_{B_1}$.
We perform 40 repetitions of the experiment and choose a subset of experiments for the qualitative evaluation.
Figure~\ref{fig:TCP_gw_reallocation} depicts the evolution of $\textnormal{\textit{STA}}_{B_1}$'s throughput during the course of two of these experiments. 
At the beginning of the experiment, a throughput of around 30 to 50~Mbps can be observed.
The fluctuations are a result of varying channel conditions and of TCP's congestion window that causes a non-steady data rate.
At the 60~seconds mark, when the gateway reallocation occurs, the end-to-end flow is interrupted for a brief time, causing a drop in the throughput.
Figure \ref{fig:TCP_gw_reallocation} captures two different cases that are representative of how the throughput evolves after the gateway reallocation process. In the first case (continuous line), TCP performs a fast recovery, quickly reaching similar throughput values as before the reallocation process.
In the second case (dashed line), TCP performs a slow start procedure, effectively extending the time it takes for the client to reach the initial throughput levels.
We conclude that, in both cases, ongoing flows using TCP recover from a gateway relocation in a reasonable time. 
Since relocation is not a common event, we prove that this is an effective mechanism to enable load-balancing while minimizing impact on ongoing traffic.
% To alleviate congestion in upper branch, SWAM access module balances Tenant B's traffic between two gateways, i.e. allocates root role at node $s4$ and $s0$ for Tenant B. traffic from STA B.1 flows through tunnel $s1$ -$s0$, whereas traffic for client STA A.1 continues to flow $s1$ - $s4$, and traffic from STA B.2 flows $s3$ - $s4$

% Note: We implement this manually (not through controller), by blocking/unblocking ports in $br\_int$ of bridges $s0$

% We plot:
% \begin{itemize}
% \item Throughput: iperf throughput between all clients and the tenant home network
% \item Latency: ping latency between STA B.1 and home network, and between STA B.1 and STA B.2. Note: We may have to run the ping tests in a separate experiment.
% \end{itemize}

% Note: We can continue the plot from the previous section. Throughput should improve because we are now balancing load through different links. Latency between STA B.1 and B.2 will change. Before link break they are one wireless hop away, after link break they are three wireless hops away, after load balancing they are two hops away but go though the tenant home network.

\subsection{Mobility}
Our last experiment showcases how SWAM deals with client mobility when clients chose to change the \emph{vap} they are connected to\footnote{In Wi-Fi the clients are in control of handover decisions}. 
For a client, the handover process should be as fast as possible in order to avoid service interruption.
In order to determine how fast SWAM can react to a client that switches from one \emph{vap} to another, two different events are analyzed: i) $\textnormal{\textit{STA}}_{A_1}$ moves from the \emph{vap} on node $s1$ to the \emph{vap} on node $s2$, and ii) $\textnormal{\textit{STA}}_{B_1}$ moves from the \emph{vap} on node $s1$ to the \emph{vap} on node $s3$ (as depicted in Fig. \ref{fig:SWAM_GWreallocation_mobility}).
It should be noted that while $\textnormal{\textit{STA}}_{A_1}$ connects to the new \emph{vap} without changing the gateway node ($s4$), $\textnormal{\textit{STA}}_{B_1}$ hands over to a \emph{vap} that has a different gateway node (from $s0$ to $s4$) to reach the tenant's home network. 
This difference is relevant, as it affects how SWAM reacts to the handover procedure, since in the second case both the access bridge in the gateway node, and the bridge in the tenant's home network need to be updated.

In order to analyze the two mobility experiments, a reachability test is performed.
It consists in performing the handover process for each client separately, while they are pinging their home network every 1~ms.
%This allows to keep track when there is connectivity and during how long the connection between the client and the tenant home network is lost when switching from one \emph{vap} to another, corresponding to the time between the last valid ICMP packet from the source \emph{vap} and the first valid ICMP packet on the destination \emph{vap}.
The overall time it takes to switch an \emph{vap} is composed of two intervals: i) the client handover time, and ii) the time for SWAM to redirect traffic to the backhaul tunnel connecting to the target \emph{vap}. 
The client handover time does not depend on SWAM and it is therefore not evaluated.
Instead, we focus our evaluation on the time required by SWAM to update the backhaul tunnels. Notice that the detailed mechanisms used to perform this update have been described in Section \ref{subsec:swam_control}.
% After the handover finishes and a client is connected to the \emph{vap}, the SWAM network reconfiguration takes place. 
% When attaching to an \emph{vap}, a client automatically generates a broadcast LLC packet that contains the client's MAC address.
% As this packet is flooded across the tenant's backhaul tunnels and it reaches the tenant's home network switch via the gateway, all the self-learning switches of the network are updated. 
% With these updates, the network reconfiguration is finished and any packet directed towards the client will use the adequate backhaul tunnels on it's way across the topology. 

%In order to evaluate the network reaction interval for $STA_{A_1}$ handover, the traffic is sniffed at the \emph{vap} interfaces of $s1$ and $s2$. 
%Fig.\ref{fig:CDF-client-handover} 
\begin{figure}
	\centering		
        \includegraphics[width=0.5\textwidth]{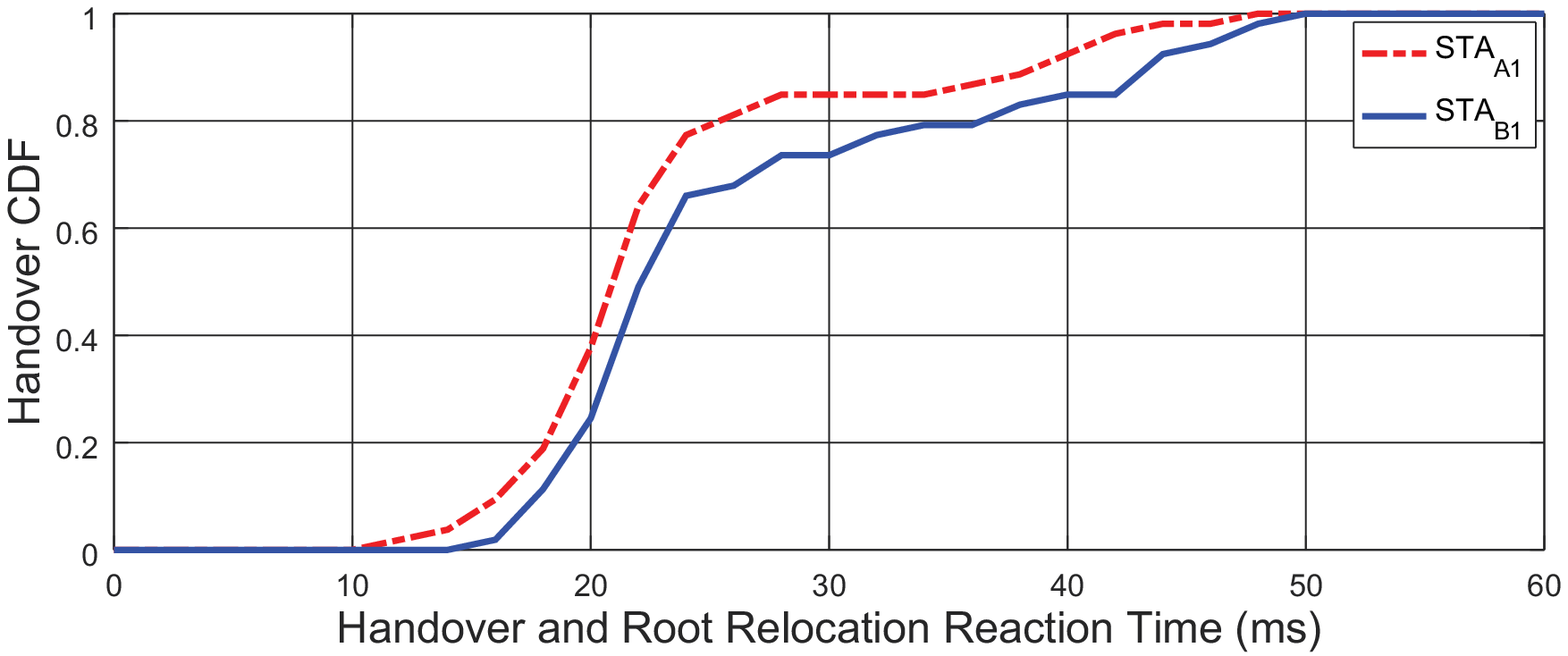}
	\caption{CDF of the network reaction time to the $\textnormal{\textit{STA}}_{A_1}$ and $\textnormal{\textit{STA}}_{B_1}$ handover.}
	\label{fig:CDF-client-handover}
\end{figure}

Figure \ref{fig:CDF-client-handover} shows the CDFs of the overall reallocation time measured for $\textnormal{\textit{STA}}_{A_1}$ and $\textnormal{\textit{STA}}_{B_1}$ across all experiment repetitions. 
The reallocation process proves to be fast, in average taking less than 30~ms and at most 50~ms, independent of the client.
In average, $\textnormal{\textit{STA}}_{A_1}$ is reallocated slightly faster than $\textnormal{\textit{STA}}_{B_1}$, as it is only required to update the access bridge in the root node, and not in the tenant's home network. %[TODO 20 ms of difference seem a lot for just reconfiguring entries of a self-learning switch. I could not find any numbers on this though (A.B)]
%[TODO is this true? MG: Yes it is, in particular the average value on 54 experiments for each station are: $AVG_{A1}$=24.12~ms, $MED_{A1}$=21.58~ms, $STD-DEV_{A1}$=8.01; $AVG_{B1}$=27.21~ms, $MED_{B1}$=23.15~ms, $STD-EV_{B1}$=9.43]
Overall, it can be stated that the reallocation times for both clients represent only a fraction of the handover time.
This demonstrates that SWAM is a very agile architecture, capable of quickly adapting its network configuration when dealing with client mobility.

%We move STA A.1 from $s1$ to $s2$, and STA B.1 from $s1$ to $s3$ as depicted in Fig.\ref{fig:SWAM_GWreallocation_mobility} (b). [TODO: Uniform to STA X.1 or $STA_{X_1}$] 
%We plot in both cases the glitch experienced in TCP downstream throughput. The difference between the two cases is that STA A.1 moves vaps without changing root node, whereas STA B.1 changes root node. We want to see the impact of changing root node. Outcome should be that TCP is not severely affected, hence streaming video would work fine.
%[TODO: Matteo can we delete the following?]
%Optional study about mobility: The quick handover is due to the spontaneous LLC-xID packet generated by the STAs when they re-attach to a new \emph{vap} (at least in Linux-based systems). We can take a bunch of STAs, e.g. our smartphone, and check if they all behave this way.

\section{Related Work}
\label{sec:related_work}

The design of the SWAM data-path is influenced by work on virtualization for multi-tenant data-centers. Especially relevant is the design of the Neutron plug-in for OpenStack \cite{openstack-neutron}. Neutron uses a hierarchical data-path composed of software switches to deliver, among other services, a layer two abstraction between virtual machines hosted in different compute nodes. The offered abstraction is equivalent to the one presented by SWAM to the virtual access points of a given tenant. Neutron uses VXLAN to establish tunnels between compute nodes over an IP infrastructure, which is a more complex tunneling scheme than the one adopted in SWAM. The H2020 5G-XHAUL project also adopts end-to-end tunnels, along with a hierarchical control plane for the transport network, to deliver virtualization and a layer two abstraction between distributed compute resources \cite{5gxhaul_slicing}. 5G-XHAUL adopts layer two tunnels based on MAC in MAC encapsulation. Tunnel paths are defined by an outer VLAN tag, and an additional inner VLAN tag is carried as a unique identifier of each virtual layer two segment, which is linked to a tenant. Like Neutron and 5G-XHAUL, SWAM also adopts end-to-end tunnels, but the data-path is simplified and tailored to smaller and bandwidth constrained outdoor Small Cells deployments. 

Control planes to provide layer two abstractions in data-centers often relay on EVPN \cite{evpn}, which uses BGP route reflectors to proactively distribute MACs between the virtualization end-points in each compute node. The main advantage of EVPN is to reduce the amount of broadcast traffic across the network by proactively populating MAC tables. Given the smaller network size and the fewer number of concurrent tenants of SWAM, as compared to modern data-center networks, we have adopted a simplified data-plane based learning, instead of the more complex distribution of MAC addresses done in EVPN. We leave for future work the study of the potential savings achieved in SWAM with centralized control plane mechanisms like EVPN. Like in SWAM, gratuitous ARPs are used in data-center overlay networks to update the location of a virtual machine after a migration event, however, to the best of our knowledge, SWAM is the first system to use them as a mobility enabler in multi-tenant Small Cell networks.

%Data-center multi-tenant overlays. Neutron using VXLAN , Netlord \cite{netlord} using Eth + IP encapsulation. 

Regarding the application of SDN to wireless mesh networks, \cite{detti} describes a hybrid architecture where a distributed OLSR daemon is used to configure the in-band control network, and a centralized OpenFlow solution is used to configure the data plane. This work demonstrates an Internet gateway balancing policy, where, unlike SWAM, the gateways are fixed and pre-established a priori. An alternative architecture based only on OpenFlow is presented in \cite{sesame_icc}, which is able to integrate point-to-multipoint (p2mp) wireless interfaces in the SDN data-path. This architecture has been adopted in SWAM to support p2mp wireless interfaces.

In the access network SWAM uses virtual access points (\emph{vaps}) to support multi-tenancy. Prior works like Odin \cite{odin} have used \emph{vaps} as primitives in Wi-Fi enterprise networks. Odin, for example, describes the concept of light virtual access points (LVAPs), which can be moved between physical APs, in order to enable load balancing and minimize the impact of client handovers. SWAM focuses on the integration of per-tenant \emph{vaps} with a wireless backhaul network, and is therefore complementary to systems that focus on the access network. 

Finally, regarding joint access-backhaul coordination schemes, the FP7 iJOIN project \cite{ijoin_access_backhaul} put forward initial architectures and identified some of the relevant challenges for dense networks of Small Cells. Similar scenarios are also being considered in the Integrated Access and Backhaul (IAB) work item of 3GPP Release 16 \cite{3gpp_iab}. Although customized for IEEE 802.11 networks, the SWAM architecture is sufficiently generic to be also applicable to 3GPP networks. The authors in \cite{senseful}, present an architecture where an access and wireless backhaul controllers, are used to allocate resources in a coordinated manner, in Small Cell networks that use IEEE 802.11 radios. This work introduces a hybrid TDMA/CSMA access scheme to enhance the coordination between access and backhaul wireless resources, which could also be adopted in SWAM.

\section{Conclusions}
\label{sec:conclusions}

In this paper we have introduced SWAM, a system that builds on inexpensive wireless devices to deliver outdoor Small Cell networks supporting multi-tenancy, client mobility, and integrated wireless access and backhaul. An infrastructure provider can integrate SWAM within an end-to-end 5G network to provide on-demand capacity slices to MNOs or vertical users. We have prototyped SWAM on an embedded platform using commercial IEEE 802.11 radios, and have demonstrated its feasibility and performance on an indoor office testbed. SWAM can recover from failures in the wireless backhaul without affecting ongoing traffic, can relocate per-tenant gateways in approximately $200$ ms, and can reconfigure backhaul tunnels upon a handover in about $20$ ms.

SWAM can be extended in several ways. First, queuing disciplines should be implemented on the physical wireless interfaces that enforce QoS on end-to-end tunnels. Second, algorithms are required to decide when to reallocate SWAM gateways in order to balance traffic.


\begin{thebibliography}{99}

\bibitem{5g_densification} 
Qualcomm, \textit{Rising to Meet the 1000x Mobile Data Challenge}. Available at: \url{https://www.qualcomm.com/media/documents/files/rising-to-meet-the-1000x-mobile-data-challenge.pdf}

\bibitem{cellular_unlicensed}
Ling, J. et al. (2015). \textit{Enhanced capacity and coverage by Wi-Fi LTE integration}. IEEE Communications Magazine, 53(3), 165-171.
%Ling, J., Kanugovi, S., Vasudevan, S., and Pramod, A. K. (2015). \textit{Enhanced capacity and coverage by Wi-Fi LTE integration}. IEEE Communications Magazine, 53(3), 165-171.

\bibitem{5G_slicing}
5GPPP Architecture Working Group, \textit{View on 5G Architecture}, June 2016. Online. Available here: \url{https://5g-ppp.eu/wp-content/uploads/2014/02/5G-PPP-5G-Architecture-WP-July-2016.pdf}

\bibitem{openflow}
The OpenFlow Switch Specification. Available at: \url{https://www.opennetworking.org}

\bibitem{openstack-neutron}
Denton, J. (2014). Learning OpenStack Networking (Neutron). Packt Publishing Ltd.

\bibitem{evpn}
Krattiger, L. et al. \textit{Building Data Centers with VXLAN BGP EVPN: A Cisco NX-OS Perspective}. Cisco Press, 2017.
%Krattiger, Lukas, Shyam Kapadia, and David Jansen. \textit{Building Data Centers with VXLAN BGP EVPN: A Cisco NX-OS Perspective}. Cisco Press, 2017.

\bibitem{sesame_icc}
Hurtado-Borras, A. et al. (2015, June). SDN wireless backhauling for Small Cells. In Communications (ICC), 2015 IEEE International Conference on (pp. 3897-3902). IEEE.
%Hurtado-Borras, A., Palà-Solé, J., Camps-Mur, D., and Sallent-Ribes, S. (2015, June). SDN wireless backhauling for Small Cells. In Communications (ICC), 2015 IEEE International Conference on (pp. 3897-3902). IEEE.

\bibitem{eucnc_16}
Betzler, A. et al. (2016, June). On the benefits of wireless SDN in networks of constrained edge devices. In Networks and Communications (EuCNC), 2016 European Conference on (pp. 37-41). IEEE.
%Betzler, A., Quer, F., Camps-Mur, D., Demirkol, I., and Garcia-Villegas, E. (2016, June). On the benefits of wireless SDN in networks of constrained edge devices. In Networks and Communications (EuCNC), 2016 European Conference on (pp. 37-41). IEEE.

\bibitem{max_vaps}
Cisco Meraki, \textit{Multi-SSID Deployment Considerations}. Available at: \url{https://documentation.meraki.com/MR/WiFi_Basics_and_Best_Practices/Multi-SSID_Deployment_Considerations}

\bibitem{5gxhaul_slicing}
Giatsios, D. et al. (2017, June). \textit{SDN implementation of slicing and fast failover in 5G transport networks}. In Networks and Communications (EuCNC), 2017 European Conference on (pp. 1-6). IEEE.
%Giatsios, D., Choumas, K., Flegkas, P., Korakis, T., and Camps-Mur, D. (2017, June). \textit{SDN implementation of slicing and fast failover in 5G transport networks}. In Networks and Communications (EuCNC), 2017 European Conference on (pp. 1-6). IEEE.

\bibitem{flrr}
SODALITE Deliverable 3: \textit{Description and initial validation of SDN algorithms, and initial experimental evaluation} Deliverable 6.10 (2015), FLEX Project, FP-7-ICT-2013-10 , Grant Agreement 612050

%\bibitem{OVS}
%The Openvswitch project. Available at: \textit{http://openvswitch.org/}

%\bibitem{ovsdb}
%Pfaff, B., and Davie, B. (2013). RFC 7047: The Open vSwitch Database Management Protocol,(2013). URL< www. ietf. org/rfc/rfc7047. txt.

\bibitem{5gxhaul}
Camps-Mur, D. et al. (2016, December). 5G-XHaul: Enabling scalable virtualization for future 5G Transport Networks. In Ubiquitous Computing and Communications and 2016 International Symposium on Cyberspace and Security (IUCC-CSS), International Conference on (pp. 173-180). IEEE.
%Camps-Mur, D., Flegkas, P., Syrivelis, D., Wei, Q., and Gutiérrez, J. (2016, December). 5G-XHaul: Enabling scalable virtualization for future 5G Transport Networks. In Ubiquitous Computing and Communications and 2016 International Symposium on Cyberspace and Security (IUCC-CSS), International Conference on (pp. 173-180). IEEE.

\bibitem{Bernardos_14}
Bernardos, C.J. et al. \textit{An architecture for software defined wireless networking,} Wireless Communications, IEEE , vol.21, no.3, pp.52,61, June 2014
%Bernardos, C.J.; De La Oliva, A; Serrano, P.; Banchs, A; Contreras, L.M.; Hao Jin; Zu\~{n}iga, J.C., \textit{An architecture for software defined wireless networking,} Wireless Communications, IEEE , vol.21, no.3, pp.52,61, June 2014

\bibitem{odin}
Lalith S. et al. \textit{Towards Programmable Enterprise WLANs With Odin,} in Proc. Workshop on Hot Topics in Software Defined Networking (HotSDN '12), 2012.
%S. Lalith, J. Schulz-Zander, R. Merz, A. Feldmann, and T. Vazao, \textit{Towards Programmable Enterprise WLANs With Odin,} in Proc. Workshop on Hot Topics in Software Defined Networking (HotSDN '12), 2012.

%\bibitem{aeroflux}
%J. Schulz-Zander, N. Sarrar, S. Schmid, (2014, August). \textit{Towards a scalable and near-sighted control plane architecture for WiFi SDNs.} In Proceedings of the third workshop on Hot topics in software defined networking (pp. 217-218). ACM.

%\bibitem{Dely_11}
%Peter Dely, Andreas Kassler, Nico Bayer. \textit{OpenFlow for Wireless Mesh Networks,} IEEE International Workshop on Wireless Mesh and Ad Hoc Networks (WiMAN 2011), Hawaii, USA, Aug. 2011.

%\bibitem{ONF_wireless}
%ONF Wireless and Mobile Working Group, available at: \textit{https://www.opennetworking.org/working-groups/wireless-mobile}

%\bibitem{LLDP}
%Sherwood, R., Gibb, G., Yap, K. K., Appenzeller, G., Casado, M., McKeown, N., and Parulkar, G. (2009). \textit{Flowvisor: A network virtualization layer}. OpenFlow Switch Consortium, Tech. Rep.

%\bibitem{MPLS}
%Luc De Ghein, \textit{MPLS Fundamentals}, Cisco Press, 2006.

%\bibitem{80211s}
%Hiertz, G. R., Denteneer, D., Max, S., Taori, R., Cardona, J., Berlemann, L., and Walke, B. (2010). \textit{IEEE 802.11 s: the WLAN mesh standard}. Wireless Communications, IEEE, 17(1), 104-111.

%\bibitem{LinuxWireless}
%Vipin, M., and Srikanth, S. (2010, January). \textit{Analysis of open source drivers for IEEE 802.11 WLANs}. In Wireless Communication and Sensor Computing, 2010. ICWCSC 2010. International Conference on (pp. 1-5). IEEE.

%\bibitem{ODL}
%OpenDayLight, available at: http://www.opendaylight.org/

%\bibitem{netlord}
%Mudigonda, J., Yalagandula, P., Mogul, J., Stiekes, B., and Pouffary, Y. (2011, August). \textit{NetLord: a scalable multi-tenant network architecture for virtualized datacenters}. In ACM SIGCOMM Computer Communication Review (Vol. 41, No. 4, pp. 62-73). ACM.

\bibitem{detti}
Detti, A. et al. \textit{Wireless mesh software defined networks (wmSDN)}. Wireless and Mobile Computing, Networking and Communications (WiMob), 2013 IEEE 9th International Conference on. IEEE, 2013
%Detti, Andrea, et al. \textit{Wireless mesh software defined networks (wmSDN)}. Wireless and Mobile Computing, Networking and Communications (WiMob), 2013 IEEE 9th International Conference on. IEEE, 2013

\bibitem{ijoin_access_backhaul}
Bernardos, C. et al. \textit{Challenges of designing jointly the backhaul and radio access network in a cloud-based mobile network.} Future Network and Mobile Summit (FutureNetworkSummit), 2013. IEEE, 2013.
%Bernardos, Carlos, et al. \textit{Challenges of designing jointly the backhaul and radio access network in a cloud-based mobile network.} Future Network and Mobile Summit (FutureNetworkSummit), 2013. IEEE, 2013.

\bibitem{3gpp_iab}
3GPP TR 38.874, \textit{NR; Study on integrated access and backhaul}, Tech.Rep., 2017.

\bibitem{senseful}
Garcia-Villegas, E. et al. (2017, June). \textit{SENSEFUL: An SDN-based joint access and backhaul coordination for Dense Wi-Fi Small Cells}. In Wireless Communications and Mobile Computing Conference (IWCMC), 2017 13th International (pp. 494-499). IEEE.
%Garcia-Villegas, E., Sesto-Castilla, D., Zehl, S., Zubow, A., Betzler, A., and Camps-Mur, D. (2017, June). \textit{SENSEFUL: An SDN-based joint access and backhaul coordination for Dense Wi-Fi Small Cells}. In Wireless Communications and Mobile Computing Conference (IWCMC), 2017 13th International (pp. 494-499). IEEE.

\end{thebibliography}
\end{document}